\documentstyle[twoside,fleqn,espcrc2,epsf]{article}

\newcommand{\be}{\begin{equation}}
\newcommand{\ee}{\end{equation}}
\newcommand{\ben}{\begin{eqnarray}}
\newcommand{\een}{\end{eqnarray}}

\title{Finite\,-Temperature QCD on the Lattice}

\author{Akira Ukawa\address{Institute of Physics, University of Tsukuba,
Tsukuba, Ibaraki 305, Japan}}

\begin{document}

\begin{abstract}
Recent developments in finite-temperature studies of lattice QCD are reviewed.
Topics include (i) tests of improved actions for the pure gauge
system, (ii) scaling study of the two-flavor chiral transition and restoration
of $U_A(1)$ symmetry with the Kogut-Susskind quark action, (iii) present
understanding of the finite-temperature phase structure for the Wilson quark
action.  New results for finite-density QCD are briefly discussed.

\end{abstract}

\maketitle
\section{Introduction}

Finite-temperature studies of lattice QCD have been pursued over a number of
years.  Quite clearly the pure gauge system is the best understood of the
entire
subject.  The system has a well-established first-order deconfinement
transition\cite{ukawa89}, and extensive and detailed results are already
available for a number of thermodynamic quantities\cite{bielefeldplaquette}.
Nonetheless many new studies have been made for this system recently.  The
purpose is to examine to what extent cutoff effects in thermodynamic quantities
are reduced for improved actions as compared to the plaquette action which had
been used almost exclusively in the past.

Full QCD thermodynamics with the Kogut-Susskind quark action has also been
investigated extensively in the past.  A basic question for this system is the
order of chiral phase transition for light quarks.  For the system with two
flavors, finite-size analyses carried out around 1989-1990 indicated an absence
of phase transition down to the quark mass
$m_q/T\approx 0.05$\cite{gottliebreview}, and a more recent
study\cite{karschlaermann} attempted to find direct evidence for the
second-order
nature of the transition, as suggested by the sigma model
analysis in the continuum\cite{sigmamodel}, through scaling analyses.
Scaling studies have been continued this year to establish the universality
nature of the transition on a firm basis.  Another issue discussed at the
Symposium is the question of restoration of
$U_A(1)$ symmetry at the chiral transition.  Results have been presented
for equation of state both without and with use of improved actions.

Studies of thermodynamics with the Wilson quark action is much
less developed compared to that for the Kogut-Susskind quark action. Past
simulations found a number of unexpected features, which made even an
understanding of the phase structure a non-trivial problem\cite{wilsonreviews}.
Recently, however, considerable light has been shed on this problem through an
analysis based on the view that the critical line of vanishing pion mass marks
the point of a second-order phase transition which spontaneously breaks parity
and flavor symmetry\cite{aoki}.
Some new work with improved actions, which was initiated a few
years\cite{qcdpaximp}, has also been made this year.

In this article we review recent studies of
finite-temperature lattice QCD.  In Sec.~2 we summarize
results for the pure gauge deconfinement transition obtained with a variety of
improved actions.  Results for the two-flavor chiral transition for the
Kogut-Susskind quark action are discussed in Sec.~3 with the main part devoted
to scaling analyses of the order of the transition.  In Sec~4 we describe
recent
progress on the phase structure analysis for full QCD with the Wilson quark
action.  This year's results for finite density are briefly discussed in Sec~5.
Our summary and conclusions are presented in Sec.~6.

\section{Recent work on pure gauge system}

\begin{table}[bt]
\setlength{\tabcolsep}{0.1pc}
\newlength{\digitwidth} \settowidth{\digitwidth}{\rm 0}
\catcode`?=\active \def?{\kern\digitwidth}
\caption{Recent work on pure gauge system with improved actions. Argument for
$\beta_c$ means spatial volume, $\mu(L)$ the torelon mass for spatial size
$L$ and $N_t$ the temporal lattice size.}
\label{tab:puregauge}
\begin{tabular}{lllc}
\hline
action & ref. &  measurements & $N_t$\\
\hline\\[-3mm]
\multicolumn{4}{l}{\underline{RG-improved}}\\
\\[-3mm]
FP (type I) & \cite{perfect1} & $\beta_c(\infty),\mu(L)$ & $2,3,4,6$\\
FP (type IIIa) & \cite{perfect3} & $\beta_c(\infty)$ & $2,3,4,6$\\
FP (type IIIa) & \cite{papa}     & $p$               & $2,3$\\
RG(1,2)        & \cite{kaneko}   & $\beta_c(\infty),\sigma$ & $4,6$\\
               & \cite{bock}     & $\beta_c(\infty),\mu(L)$ & $ 2,3$\\
\\[-2mm]
\multicolumn{4}{l}{\underline{Symanzik-improved}}\\
\\[-3mm]
S(1,2)$_{\mbox{\small tree}}$   & \cite{cella}    & $\beta_c(4N_t)$ & $
3,4,5,6$\\
               & \cite{bielefeld12}&
$\beta_c(\infty),\sigma,\epsilon,p,\sigma_I$
& $4$\\
S(1,2)$_{\mbox{\small tadpole}}$ & \cite{bielefeld12}&
$\beta_c(\infty),\sigma,\epsilon,p,\sigma_I$ &
$4$\\
S(2,2)$_{\mbox{\small tree}}$ & \cite{bielefeld22,bielefeld12}&
$\beta_c(\infty),\sigma,\epsilon,p$ & $4$\\
SLW$_{\mbox{\small tadpole}}$ & \cite{cornell} & $\beta_c(2N_t),\mu(L)$ &
$2,3,4$\\
            & \cite{bock}    & $\beta_c(\infty),\mu(L)$ & $2,3$ \\
\hline
\end{tabular}
\vspace{-8mm}
\end{table}

In Table~\ref{tab:puregauge} we list recent studies of the pure gauge system
using improved actions.
Among actions constructed through renormalization group,
RG(1,2)\cite{iwasakirg} includes $1\times 2$ loop in addition to the
plaquette.  The action FP is an 8-parameter approximation to the fixed point
action\cite{perfect1} containing plaquette and unit parallelogram and
their power up to 4;  type I\ \cite{perfect1} and type IIIa\ \cite{perfect3}
differ in the blocking procedure, the latter including 5 and 7 link staples to
improve rotational symmetry. Actions improved according to Symanzik's
program with the addition of an $m\times n$ loop are denoted as S(m,n),
distinguishing tree-level coefficients from tadpole-improved
ones\cite{lepagemackensie} by corresponding suffices.  The S(1,2) action
improved to one loop order\cite{luescherweisz} including tadpole
factors\cite{lswtadpole} is denoted by SLW$_{\mbox{\small tadpole}}$.

\subsection{Critical temperature}

\subsubsection{$T_c/\protect\sqrt{\sigma}$}

\begin{figure}[t]
\centerline{\epsfxsize=74mm \epsfbox{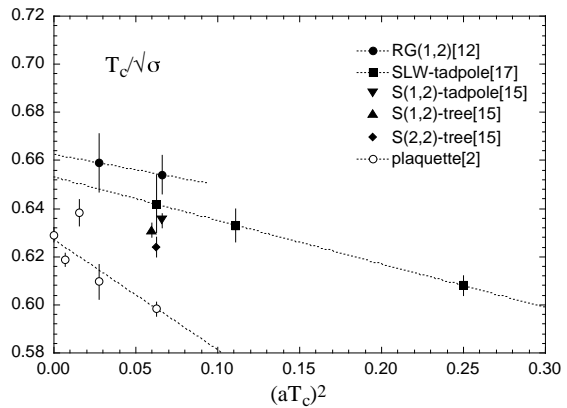}}
\vspace{-8mm}
\caption{$T_c$ for pure gauge system normalized by
square root of string tension $\protect\sqrt{\sigma}$ as a function of
$(aT_c)^2$.  Dotted lines are
$O(a^2)$ fits.}
\vspace*{-8mm}
\label{fig:fig1}
\end{figure}

A basic quantity for the pure gauge system is the ratio
$T_c/\sqrt{\sigma}$, where $\sigma$ is the string tension extracted from the
static quark-antiquark potential.  In Fig.~\ref{fig:fig1} we plot recent
results
for this ratio for improved actions and the plaquette
action\cite{bielefeldplaquette} as a function of $(aT_c)^2=1/N_t^2$ with $N_t$
the temporal lattice size.  Results are not available for fixed point actions.
For SLW$_{\mbox{\small tadpole}}$, for which values of $\sigma$ are not given
by
the authors\cite{cornell}, we use an interpolation of the results reported in
ref.~\cite{lswscri}.

Scaling behavior can be examined for two types of improved actions,
SLW$_{\mbox{\small tadpole}}$ and RG(1,2).  Comparing their results with those
of
the plaquette action, it is apparent that the improved actions exhibit a better
scaling behavior.  For SLW$_{\mbox{\small tadpole}}$ the ratio increases by
about
5\% over $aT_c\approx 0.25-0.5$,  while a variation of $3-7$\% occurs for
the plaquette action over a factor two smaller range $aT_c\approx 0.08-0.25$.
The results for RG(1,2) are constant over $aT_c\approx 0.17-0.25$ within the
quoted error of $1-2$\%. The errors are still sizable, however, to draw a
conclusion on the magnitude of slope from the limited range of lattice spacing
explored so far.

We observe, however, that the continuum extrapolation of $T_c/\sqrt{\sigma}$
for the improved actions do not agree with that for the plaquette action.   For
SLW$_{\mbox{\small tadpole}}$, assuming an $O(a^4)$ or
$O(a^2)$ dependence (we ignore the factor $g^4$ of the actual form $O(g^4a^2)$)
both of which are consistent with present data, we find
$T_c/\sqrt{\sigma}=0.641(7)$ and 0.653(10) in the continuum limit.  For
RG(1,2) we expect an $O(a^2)$ scaling violation, with which we obtain
$T_c/\sqrt{\sigma}=0.663(13)$.  These values are $2-5$\% (one to two standard
deviations) larger compared to the estimate for the  plaquette action
$T_c/\sqrt{\sigma}=0.629(3)$\cite{bielefeldplaquette} (open circle at
$aT_c=0$) obtained with a quadratic extrapolation.

We stress that efforts to resolve the discrepancy should be made for we would
then have a determination of the basic ratio $T_c/\sqrt{\sigma}$ accurate at
the level of $1-2$\%.

Let us add a remark on the results at $aT_c=0.25$ where values for six
types of actions are available.  Among those belonging to the category of
Symanzik improvement, we observe that $T_c/\sqrt{\sigma}$ systematically
increases in the order,  plaquette$\to$S(1,2)$_{\mbox{\small tree}}$
$\to$S(1,2)$_{\mbox{\small tadpole}}$ $\to$ SLW$_{\mbox{\small tadpole}}$,
reaching $T_c/\sqrt{\sigma}\approx 0.64$.  If we take $T_c/\sqrt{\sigma}\approx
0.65$ as the continuum value(see above), this trend is consistent with
the theoretical expectation that cutoff effects are reduced with an
increasing degree of improvement. In this regard  S(2,2)$_{\mbox{\small tree}}$
seems
less improved than S(1,2)$_{\mbox{\small tree}}$.  The fact that the value for
RG(1,2) lies above that for SLW$_{\mbox{\small tadpole}}$ may be ascribed to
a larger coefficient of the $1\times 2$ loop term for this action compared to
that for the latter.

\subsubsection{$T_c/\protect\sqrt{\sigma(L)}$}
\label{sec:torelon}

\begin{figure}[t]
\centerline{\epsfxsize=74mm \epsfbox{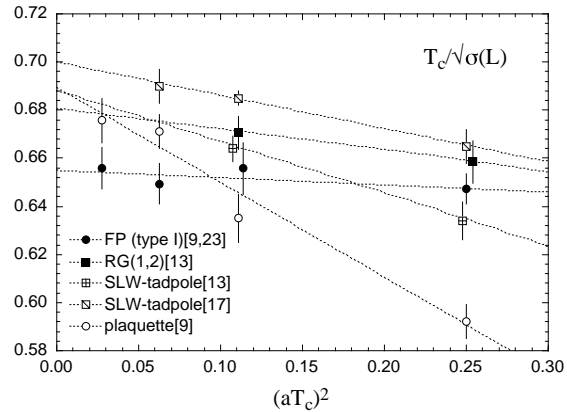}}
\vspace{-8mm}
\caption{$T_c$ normalized by $\protect\sqrt{\sigma(L)}$ as a function of
$(aT_c)^2$ where $\sigma(L)=\mu(L)/L$ with $\mu(L)$ the torelon mass for a
physical spatial size $L$ fixed at
$L=2/T_c$. Dotted lines are $O(a^2)$ fits.}
\vspace*{-8mm}
\label{fig:fig2}
\end{figure}

Another quantity often used for testing improvement with simulations on small
lattices is the torelon mass $\mu (L)$ extracted from the
Polyakov loop correlator on a lattice of spatial size $L$. Defining
$\sigma(L)=\mu(L)/L$ we compile in Fig.~\ref{fig:fig2} results for the ratio
$T_c/\sqrt{\sigma(L)}$ for a fixed physical spatial size $L=2/T_c$
as a function of $(aT_c)^2$.
The results for the fixed point action FP(type I) and the plaquette action
were already available last year\cite{perfect1} except for the value for the
former for $N_t=6$\cite{degrand}. This year, results have been reported for
the actions SLW$_{\mbox{\small tadpole}}$ and RG(1,2).

Taking all the data together, one finds a better scaling behavior exhibited by
the improved actions compared to the plaquette action.  On a closer look,
however, some discrepancy is observed among data for improved actions.  While
results for FP(type I)\cite{perfect1} show a constant behavior in $aT_c$ within
the quoted error of
$1-2$\%, those for  SLW$_{\mbox{\small tadpole}}$ obtained by the Cornell
group\cite{cornell} exhibits an increase of 4\% over  $aT_c=0.5-0.25$, and
increasingly deviate from those of FP (type I) toward smaller lattice spacings.
The results for RG(1,2) lie in between those of the two actions.  There is also
a discrepancy among results from ref.~\cite{cornell} and \cite{bock}, both for
SLW$_{\mbox{\small tadpole}}$.

In order to extract the torelon mass, the Cornell group\cite{cornell} employs a
multi-state fit including excited states to a set of Polyakov loop correlators
unsmeared or smeared at either source or sink, while a single state fit to a
correlator smeared at source and sink is used in the other
studies\cite{perfect1,bock}. The discrepancy possibly
originates from contamination of excited states in the latter
fit\cite{cornell}.

Because of the different trend in the lattice spacing dependence,
continuum extrapolation leads to a significant scatter of the ratio among the
actions. Dotted lines in Fig.~\ref{fig:fig2} illustrate the difference, where
we
assume
$O(a^2)$ dependence since $O(a^2)$ terms are expected to be present for all the
actions employed.  Compared to the value for the plaquette action, the
values for SLW$_{\mbox{\small tadpole}}$ and RG(1,2) deviate at a  $\pm
(1-2)$\% level, while that for FP(type I) is 6\% lower.

The discrepancy should be resolved, particularly to ascertain if cutoff effects
are reduced to within $1-2$\% at a large lattice spacing of
$aT_c=0.25-0.5$ for the fixed point action.

Another problem concerns the consistency between the values of
$T_c/\sqrt{\sigma(L)}$ and $T_c/\sqrt{\sigma}$.  The relation
$\sigma\approx \sigma(L)+\pi/(3L^2)$ which holds in string
models\cite{forcrand}
suggests that $T_c/\sqrt{\sigma(L)}$ increases by about 0.04 compared to
$T_c/\sqrt{\sigma}$ for $L=2/T_c$.  Making a comparison for each action, we
find that the increase for SLW$_{\mbox{\small tadpole}}$ is consistent with
this
estimate. For the plaquette action the value of $T_c/\sqrt{\sigma(L)}$ in the
continuum limit is 3\% higher than the estimate, while for RG(1,2) the value
is 3\% lower.

A systematic comparative analysis of various actions employing the same
simulation and analysis procedures is need to resolve this discrepancy.

\subsection{Latent heat and interface tension}

\begin{figure}[t]
\centerline{\epsfxsize=74mm \epsfbox{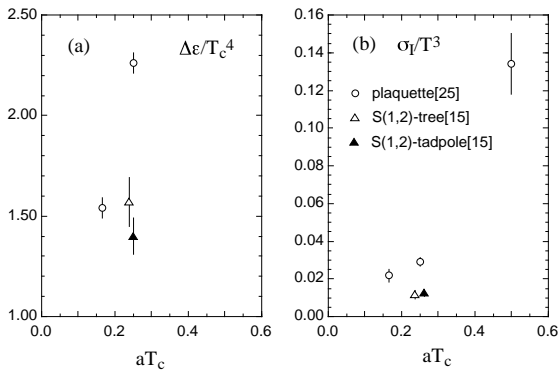}}
\vspace{-8mm}
\caption{(a) latent heat $\Delta\epsilon/T_c^4$ and (b) interface tension
$\sigma_I/T_c^3$ for plaquette\protect\cite{latentheat} and Symanzik-improved
S(1,2) action\protect\cite{bielefeld12}.}
\vspace*{-8mm}
\label{fig:fig3}
\end{figure}

The latent heat $\Delta\epsilon$ and interface tension $\sigma_I$ at the
deconfinement transition exhibit large scaling violations from an $N_t=4$ to
an $N_t=6$ lattice for the plaquette action as shown by open circles in
Fig.~\ref{fig:fig3}. The Bielefeld group measured these quantities for the
S(1,2)
action\cite{bielefeld12} both without and with tadpole improvement(triangles).

For both quantities the rapid decrease of values for the plaquette
action indicates that the continuum value would be lower than that on an
$N_t=6$ lattice.  The values for the improved actions are
indeed smaller already for $N_t=4$.

\subsection{Energy density and pressure}

Bulk thermodynamic quantities such as energy density and pressure receive
substantial contributions from high momentum modes.  Since cutoff effects in
these fluctuations are directly reduced by the improvement procedure, we may
expect significant improvement.

Numerical tests have been made\cite{bielefeld12,bielefeld22} for the actions
S(1,2)$_{\mbox{\small tree,\ tadpole}}$ and S(2,2)$_{\mbox{\small tree}}$
through comparison of pressure calculated on an $N_t=4$ lattice with an
estimate
of the continuum value obtained with the plaquette
action\cite{bielefeldplaquette}.  The results for the improved actions are
close
to the continuum estimate already for $N_t=4$,  especially for
S(1,2)$_{\mbox{\small tadpole}}$ for which a good agreement is seen even close
to
$T_c$.  The fixed point action FP(type IIIa) also exhibits a similar agreement
for $N_t=3$\cite{papa}.  We refer to Figure 1 of ref.~\cite{bielefeld12} for
these points.

\subsection{Summary}

Improved actions lead to a sizable reduction of cutoff effects in
thermodynamic quantities.  Detailed scaling studies, however, are
limited to that of critical temperature, for which consistency of
results for various actions are not yet attained beyond the level of 5\%.
Further studies are needed to see if accurate determination of
thermodynamic quantities at a few percent level is possible through simulations
with a moderately large temporal size of $4\leq N_t\leq 8$ with improved
actions.

\section{Chiral phase transition with the Kogut-Susskind quark action}

\subsection{Order of transition for $N_f=2$}

\begin{table}[b]
\vspace{-5mm}
\setlength{\tabcolsep}{0.2pc}
\catcode`?=\active \def?{\kern\digitwidth}
\caption{Studies of order of $N_f=2$ chiral transition on an $N_t=4$ lattice.}
\label{tab:nf2studies}
\begin{tabular}{llll}
\hline
 &ref. &\ \ size &\ \ $m_q$\\
\hline\\[-3mm]
KEK(1990)&\cite{founf2} & $6^3,8^3,12^3$ & $0.0125, 0.025$\\
Columbia(1990)&\cite{columbianf2} & $16^3$ & $0.01, 0.025$\\
Bielefeld(1994)&\cite{karschlaermann} & $8^3$ & $0.02-0.075$\\
Bielefeld(1996)&\cite{bielefeldchiral} & $12^3,16^3$ & $0.02-0.075$\\
JLQCD(1996)    &\cite{jlqcd} & $8^3,12^3,16^3$ & $0.01-0.075$\\
\hline
\end{tabular}
\end{table}

In Table~\ref{tab:nf2studies} we list major studies on the order of two-flavor
chiral phase transition carried out on an  $N_t=4$ lattice with the spatial
size $L^3$.  In the finite-size scaling study pursued around
1989-1990\cite{founf2,columbianf2}, measurements were made of the
susceptibility
of the Polyakov line $\Omega$ given by
\be
\chi_\Omega=L^3\left[\langle (\mbox{Re}\Omega)^2\rangle-
\langle \mbox{Re}\Omega\rangle^2\right]
\ee
and a pseudo-susceptibility of chiral order parameter defined by
\be
\chi_c=\frac{1}{9L^3}\left[
\left\langle\left(\xi^\dagger D^{-1}\xi\right)^2\right\rangle-
\left\langle\xi^\dagger D^{-1}\xi\right\rangle^2\right]
\ee
where $\xi$ is a gaussian noise and $D$ the Kogut-Susskind quark operator.
It was found that the peak height of the two susceptibilities increases up to
$12^3$, but stays constant within errors between $12^3$ and
$16^3$ both at $m_q=0.025$ and $0.0125-0.01$ (see Fig.~11 and 12 in the second
paper of  ref.~\cite{columbianf2}).  The saturation implies the absence of
a phase transition down to
$m_q\approx 0.01$. Since this quark mass is quite small, corresponding to
$m_\pi/m_\rho\approx 0.2$ at the point of the transition $\beta_c\approx 5.27$,
it was thought that the result is consistent with the transition being of
second
order at $m_q=0$ as suggested by the sigma model analysis\cite{sigmamodel}.

\subsubsection{Scaling analysis of susceptibilities}

One can attempt to examine if the transition is of second order employing
the method of scaling analysis.  Let us define the susceptibilities $\chi_m$
and
$\chi_{t, i}$\ $(i=f, \sigma, \tau)$ by
\ben
\chi_m&=&V\left[\langle (\overline{q}q)^2\rangle-
\langle\overline{q}q\rangle^2\right]\\
\chi_{t,f}&=&V\left[\langle \overline{q}q\cdot\overline{q}D_0q\rangle-
\langle\overline{q}q\rangle\langle\overline{q}D_0q\rangle \right]\\
\chi_{t,i}&=&V\left[\langle \overline{q}q\cdot P_i\rangle-
\langle\overline{q}q\rangle\langle P_i\rangle \right], \quad i=\sigma,\tau
\een
with $V= L^3N_t$, $D_0$ the temporal component of the Dirac operator, and
$P_{\sigma, \tau}$ the spatial and temporal plaquette.   For a given quark mass
$m_q$, let
$g^{-2}_c(m_q)$ be the peak position of
$\chi_m$ as a function of the coupling constant $g^{-2}$, and let
$\chi_m^{max}$ and $\chi_{t,i}^{max} (i=f, \sigma, \tau)$ be the peak height.
For a second-order transition, these quantities are expected to scale toward
$m_q\to 0$ as
\ben
g^{-2}_c(m_q)&=&g^{-2}_c(0)+c_gm_q^{z_g}\\
\chi_m^{max}&=&c_mm_q^{-z_m}\\
\chi_{t,i}^{max}&=&c_{t,i}\,m_q^{-z_{t,i}},\quad i=f,\sigma,\tau
\label{eq:exponents}
\een
Let us note that $\chi_{t, i}$\ $(i=f, \sigma, \tau)$ are three
parts of the susceptibility
$
\chi_{t}=V\left[\langle \overline{q}q\cdot \epsilon\rangle-
\langle\overline{q}q\rangle\langle \epsilon\rangle \right]
$
with $\epsilon$ the energy density\cite{karschlaermann}.  The leading exponent
is therefore given by $z_t=\mbox{Max}(z_{t,f}, z_{t,\sigma}, z_{t,\tau})$.

Natural values to expect for the exponents $z_g, z_m$ and
$z_t$ at a finite lattice spacing are those of $O(2)\approx U(1)$ corresponding
to the exact symmetry group of the Kogut-Susskind action. However, sufficiently
close to the continuum limit where flavor breaking effects are expected to
disappear,  they may take the values for $O(4)\approx SU(2)\otimes SU(2)$ which
is the group of chiral symmetry for $N_f=2$ in the continuum. One should also
remember that mean-field exponents control the scaling behavior not too close
to
the transition.  A possibility of mean-field exponents arbitrarily close to the
critical point has also been discussed\cite{kocickogut}.

The initial scaling study was carried out by Karsch and
Laermann\cite{karschlaermann} employing an
$8^3\times 4$ lattice and $m_q=0.02, 0.0375, 0.075$.  Compared to the
$O(4)$ values their results for exponents show a good agreement of $z_m$, a
50\% larger value for $z_g$ and a value twice larger for $z_t$.  Comparison
with $O(2)$ and mean-field exponents is similar since they are not too
different from the $O(4)$ values.

This work had limitations in several respects: (i) the scaling formulae are
valid
for a spatial size large enough compared to the correlation length.  At
$m_q=0.02$ the pion correlation length equals
$\xi_\pi\approx 3$.  Whether the spatial size of $L=8$ employed
is sufficiently large has to be examined.  (ii) The size of
the scaling region in terms of quark mass is {\it a priori} not known.
Hence the behavior for smaller quark masses should be explored to
check if the results are not affected by sub-leading and
analytic terms in an expansion of susceptibilities in $m_q$.  (iii)
In the original work the noisy estimator with a single noise
vector was employed to estimate disconnected double quark loop
contributions.  This introduces contamination from connected diagrams and local
contact terms, which has to be removed. Other factors such as step size of the
hybrid R algorithm and stopping condition for the solver of Kogut-Susskind
matrix could also affect the value of susceptibilities.

For these reasons the Bielefeld group has continued their
study\cite{bielefeldchiral}, and the JLQCD Collaboration\cite{jlqcd} has
started
their own work last year.  As one sees in Table~\ref{tab:nf2studies} run
parameters of new simulations are chosen to examine the points (i) and (ii)
above.  In order to deal with (iii) Bielefeld group worked out the correction
formula for the case of the single noise vector.  They also employed the
method of multiple noise vectors for some of the runs.  JLQCD employed the
method of wall source  without gauge fixing\cite{wall}, and removed
contamination by a correction formula.  At present both groups have
accumulated $(5-10)\times 10^3$ trajectories of unit length with a small step
size of $\delta\tau=(1-1/2)m_q$ for each value of
$\beta$ and $m_q$.  The standard reweighting
technique\cite{reweighting} is used to find the peak of susceptibilities.

\begin{figure}[t]
\centerline{\epsfxsize=74mm \epsfbox{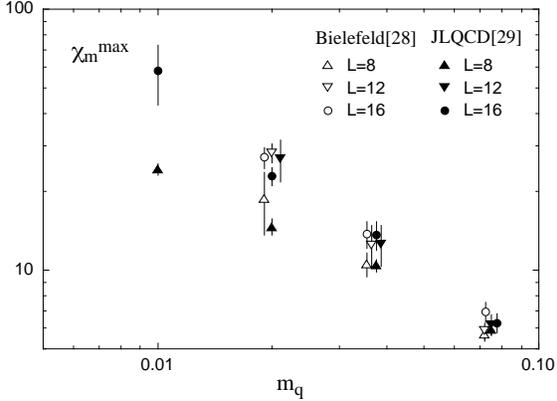}}
\vspace{-8mm}
\caption{Peak height of $\chi_m$ as a function of
$m_q$ for spatial sizes $L=8,12,16$.}
\vspace*{-8mm}
\label{fig:fig4}
\end{figure}

In Fig.~\ref{fig:fig4} we plot results of the two groups for the peak
height of the chiral susceptibility $\chi_m$ for three spatial sizes
$L=8, 12$ and 16.  The two results are consistent.  A striking feature in
Fig.~\ref{fig:fig4} is that, except for the largest quark mass
$m_q=0.075$, the peak height exhibits a significant size dependence whose
magnitude rapidly increases toward small quark masses.  In more detail, for
$m_q=0.0375$ and 0.02, the increase is largest between $L=8$ and 12, while the
peak height is consistent between $L=12$ and 16.  At $m_q=0.01$ the peak
height increases by a factor 3 between $L=8$ and 16.  Data on a $12^3$
lattice is not yet available for this quark mass.

We should remark that the new data for $m_q=0.01$ for $L=16$\cite{jlqcd} do not
agree with the previous results reported in ref.~\cite{columbianf2}.  For the
susceptibilities $\chi_c$ and $\chi_\Omega$ for which a direct comparison is
possible, new simulations give values which are a factor two larger than the
old
results.  It is possible that a smaller statistics (2500
trajectories\cite{columbianf2} as compared to 4400\cite{jlqcd}), and perhaps
also a slightly smaller estimate of the critical coupling
($\beta_c=5.265$ as compared to $\beta_c=5.266$), led to an underestimate of
susceptibilities in ref.~\cite{columbianf2}.

\begin{table}[bt]
\setlength{\tabcolsep}{0.2pc}
\catcode`?=\active \def?{\kern\digitwidth}
\caption{Exponents for $N_f=2$.  For each exponent first row represents
Bielefeld results\protect\cite{bielefeldchiral} and second those of
JLQCD\protect\cite{jlqcd}. Results are not yet available for entries marked
with
\lq\lq$-$\rq\rq. }
\label{tab:newexponents}
\vspace{1mm}
\begin{tabular}{lllllll}
\hline
& $O(2)$ & $O(4)$ & MF &$L=$8&$L=$12&$L=$16\\
\hline\\[-3mm]
$z_g$&0.60&0.54&2/3&0.77(14)&$\quad -$&$\quad -$\\
     &    &    &   &0.70(10)&$\quad -$&0.63(5)\\
\hline
$z_m$&0.79&0.79&2/3&0.79(4)&1.05(8)&0.93(9)\\
     &    &    &   &0.70(3)&1.01(11)&1.02(7)\\
\hline
$z_t$&0.39&0.33&1/3&\\
\\[-3mm]
$z_{t,f}$&&&       &0.65(7)&$\quad -$&$\quad -$\\
     &    &    &   &0.42(4)&0.75(12)&0.78(8)\\
$z_{t,\sigma}$&&&  &0.63(7)&0.96(12)&0.86(11)\\
     &    &    &   &0.48(4)&0.79(14)&0.81(9)\\
$z_{t,\tau}$&&&    &0.63(7)&0.94(13)&0.85(12)\\
     &    &    &   &0.47(4)&0.81(14)&0.82(9)\\
\hline
\end{tabular}
\vspace*{-5mm}
\end{table}

We list the exponents obtained through fits of form
(6-8) in Table~\ref{tab:newexponents}.  While some
systematic difference appears present for $z_{t,i}$ between the two groups, a
significant increase of $z_m$ and  $z_{t,i}$ from
$L=8$ to $L=12-16$ is evident, with the values for larger sizes sizably
deviating
from either $O(4), O(2)$ or the mean-field predictions.  For $z_g$ the
deviation seems less apparent though full data are not yet available.

A puzzling nature of the values of exponents becomes clearer if we translate
them into the more basic thermal and magnetic exponents $y_t$ and $y_h$
using the relations,
\be
z_g=\frac{y_t}{y_h},\ z_m=\frac{d}{y_h},\ z_t=\frac{y_t}{y_h}+\frac{d}{y_h}-1
\label{eq:rgexponents}
\ee
with $d=3$ the space dimension.  The values in Table~\ref{tab:newexponents}
are reasonably consistent with the relation $z_g+z_m=z_t+1$ which follow from
(\ref{eq:rgexponents}).  We observe that $z_m\approx 1.0(1)$ obtained for
larger spatial lattices implies
$y_h\approx 3.0(3)$ to be compared with the $O(4)$ value 2.49, while
$y_h=d=3$ is expected for a first-order phase transition.  For the
thermal exponent we find $y_t\approx 2.4(3)$ if we take $z_t\approx 0.8(1)$
or $y_t\approx 2.7(3)$ for $z_t\approx 0.9(1)$,
which is substantially larger than the $O(4)$ value of 1.34.

One may think of various possibilities for the reason leading to these
values of exponents. \linebreak (i)  The most conventional would be that
the influence of sub-leading and analytic terms is still sizable at the range
of
quark mass explored.  (ii) Another possibility, suggested by the value
$y_h\approx d$ for $L=12$ and 16, is that a discontinuity fixed  point with
$y_h=d$\cite{discontinuity} controlling the first-order transition along the
line $m_q=0$ in the low-temperature phase is strongly influencing the scaling
behavior.  The transition is of second order in this case. Whether the
deviation
of $y_t$ from any of the expected values can be explained is not clear,
however.  (iii) The transition is of second order with the exponents close to
but not equal to $d$.  This would mean a significant departure from the
universality concepts, stepping even beyond the suggestion of mean-field
exponents arbitrarily close to the critical point\cite{kocickogut}.  (iv)
The transition is of first order. In this case,
the value of quark mass $m_q=m_q^c$ at which the first-order transition
terminates  would have to be small or even vanish since the scaling formula is
derived under the assumption of a transition taking place at a single point at
$m_q=0$.

Concerning the possibility (iv), results of present data examined from
finite-size scaling point of view are as follows.  As we already pointed out,
$\chi_m$ for a fixed value of $m_q$ stays constant for
$L=12-16$ down to $m_q=0.02$.  Results for other susceptibilities
exhibit a similar behavior.  Thus a phase transition does not exist for
$m_q\geq
0.02$ as concluded in the previous studies\cite{founf2,columbianf2}.
At $m_q=0.01$ the susceptibilities increase by a factor
3 between $L=8$ and 16. Runs for $L=12$ are needed to see if the
increase is consistent with a linear behavior in volume expected for a
first-order transition.

We have to conclude that scaling analyses of susceptibilities carried out so
far do not allow a definite conclusion.  Much further work, possibly with a
quark
mass smaller than has been explored so far, is needed to elucidate the nature
of
the chiral transition for $N_f=2$.

\subsubsection{Scaling analysis of chiral order parameter}

For a second-order transition the singular part of the
chiral order parameter is expected to scale as
$\langle\overline{q}q\rangle=m_q^{1/\delta}\phi\left(x\right)$
where $\phi(x)$ is a function of the scaling variable
$x=(g^{-2}-g^{-2}_c(0))/m_q^{1/\beta\delta}$, and $1/\delta=1-d/y_h$ and
$1/\beta\delta=y_h/y_t$.  A previous analysis\cite{detarreview} employing a
collection of data generated over the years did not find clear sign of scaling
with the $O(4)$ exponents.  This year the MILC Collaboration attempted a more
elaborate analysis as part of their study of equation of state\cite{milcks96}.

For systems in the $O(4)$ universality class, the scaling function
$\phi(x)$ may be determined in a parameterized form through a simulation of
the $O(4)$ sigma model up to an overall constant for the
scaling variable $x$\cite{toussaint}. The result for $\phi(x)$ is used to fit
data for
$\langle\overline{q}q\rangle$ generated on a $12^3\times 6$ lattice with
$m_q=0.025$ and 0.0125.  Adding an analytic term of form
$m_q(c_0+c_1/g^2+c_2/g^4)$, the fit was found acceptable for $O(4)$ and also
for
the mean-field scaling function.  Extrapolating to the limit $m_q=0$, the
results differ significantly between the two cases, however (see Fig.~2 of
ref.~\cite{milcks96}).

We note that the results of the present MILC analysis do not contradict those
of
susceptibilities: the quark mass used for this work corresponds to
$m_q\approx 0.02-0.04$ on an $8^3\times 4$ lattice, for which case the
exponents found from susceptibilities are similar to the $O(4)$ values.
We further remind, however, that the exponents exhibit a significant size
dependence.  This means that studies with larger lattice sizes and smaller
$m_q$ are required to explore the nature of the two-flavor transition
from scaling of the chiral order parameter.

\subsection{Restoration of $U_A(1)$ symmetry}

For sufficiently high temperatures topologically
non-trivial gauge configurations are suppressed, leading to restoration of
$U_A(1)$ symmetry.  To what extent $U_A(1)$ symmetry is restored close to the
chiral transition is an interesting question.

\begin{figure}[t]
\centerline{\epsfxsize=74mm \epsfbox{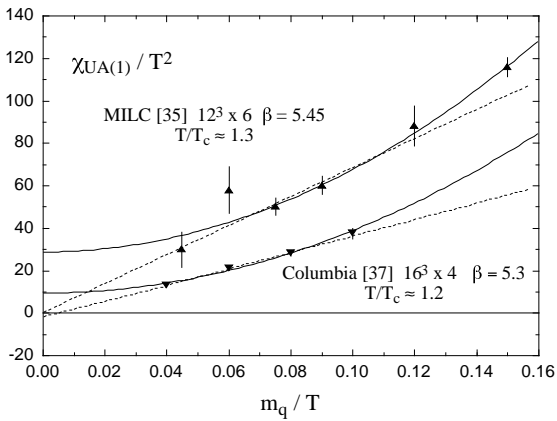}}
\vspace{-8mm}
\caption{Susceptibility for $U_A(1)$ symmetry as a function of $m_q/T$.}
\vspace*{-8mm}
\label{fig:fig5}
\end{figure}

Three groups\cite{milcks96,columbiaanomaly,bielefeldchiral}
examined the problem using the susceptibility defined by
\be
\chi_{U_A(1)}=\int d^4x\left(\langle\vec\pi(x)\cdot\vec\pi(0)\rangle-
\langle\vec a_0(x)\cdot\vec a_0(0)\rangle\right)
\ee
which should vanish at $m_q=0$ if $U_A(1)$ symmetry is restored.
In Fig.~\ref{fig:fig5} we plot the $m_q$ dependence of this quantity obtained
by
the MILC Collaboration\cite{milcks96} and the
Columbia group\cite{columbiaanomaly}.
Both results are taken in the high temperature phase corresponding
to $T/T_c\approx 1.2-1.3$. While the data appear to extrapolate linearly to
zero
at
$m_q=0$ (dotted lines)\cite{columbiaanomaly}, it is more reasonable to fit with
a
quadratic dependence
$\chi_{U_A(1)}=a+bm_q^2+O(m_q^4)$ (solid lines)\cite{milcks96} since the
susceptibility is expected to be an analytic function of $m_q$ and hence even
in $m_q$ in the high temperature phase.  These fits, which have reasonable
$\chi^2$, lead to a non-zero value of
$\chi_{U_A(1)}$ at $m_q=0$.  Thus effect of anomaly still breaks
$U_A(1)$ symmetry just above the chiral transition.   Similar results were
reported in ref.~\cite{bielefeldchiral}.

The Illinois group\cite{illinois} calculated the screening mass of $\sigma,
\pi, a_0$ on a $16^3\times 8$ lattice at $m_q=0.00625$, employing a noisy
estimator for the disconnected contribution for the $\sigma$ propagator. They
found a decrease of the mass splitting $m_\pi-m_{a_0}$ across the transition,
which, however, remains at the level of 20\% just above $T_c$ for the quark
mass employed.  They have also shown that the disconnected part of the
$\sigma$ propagator is dominated by fermionic modes with small eigenvalues
induced by instantons.  The latter results parallel those of a previous study
carried out for the quenched Wilson case\cite{iwasakianomaly}.

\subsection{Energy density and pressure}

The MILC Collaboration completed their study of equation of
state for the temporal size $N_t=6$\cite{milcks96}, following their previous
work for $N_t=4$\cite{milceos94}.  They also attempted an
extrapolation of equation of state toward $m_q\to 0$ making use of the scaling
function computed for the chiral order parameter.

The Bielefeld group\cite{bielefeld12} made a measurement of energy density
for $N_f=4$ employing the three-link improved form of the Kogut-Susskind
action\cite{naik} together with S(1,2)$_{\mbox{\small tree}}$ for the gauge
action.

In the results of both groups the energy density $\epsilon/T^4$ rapidly
rises across $T_c$ and stays close to the continuum Stefan-Boltzmann value
in the high temperature phase. In the MILC result a bump just
above $T_c$ seen for $N_t=4$ almost disappears for $N_t=6$. A similar bump,
while observed by the Bielefeld group at finite $m_q$ in their $N_f=4$ results,
is no longer present if terms vanishing in the chiral limit is removed.

These results indicate that $\epsilon/T^4$ for full QCD is a monotonically
increasing function of temperature similar in shape to that of the pure
gauge system\cite{bielefeldplaquette}.  The behavior of pressure $3p/T^4$ is
also
similar, rising smoothly from $T\approx T_c$ and reaching $\epsilon/T^4$ at
$T/T_c\approx 2-3$.

\section{Phase structure for the Wilson quark action}

\subsection{Previous phase diagram studies}
\label{sec:wilsonone}

A basic concept in the phase structure analysis for the Wilson quark action
is that of the critical line $K=K_c(\beta)$ which is usually defined as the
line of vanishing pion mass. At zero temperature this line runs from $K\approx
1/4$ at $\beta=0$ to $K=1/8$ at
$\beta=\infty$, and chiral symmetry is expected to become restored toward the
weak-coupling limit along this line\cite{bocchichio}.

At finite temperatures there also exists the line of finite-temperature
transition $K=K_t(\beta)$, the thermal line.  This line starts from the
point of the deconfinement transition of the pure gauge system at $K=0$, and
moves toward the critical line. Naively one would expect the thermal line to
hit
the critical line at some finite $\beta=\beta_{ct}$, separating the physical
region $K\leq K_c(\beta)$ into low and high temperature phases.

Extensive studies have been carried out to examine if this expectation is
realized\cite{fouwil,ukawawil,guptawil,bitarwil,hemcgcwil,qcdpaxwil,milcwil}.
A conceptual issue that arose in the course of studies is whether one can
naturally define the critical line in the high temperature phase since pion
mass does not vanish in this phase.

The QCDPAX Collaboration took the
view\cite{qcdpaxwil} that the critical line should be defined by the
vanishing
of the quark mass
$m_q$ at zero temperature, where $m_q$ is defined through chiral Ward
identity\cite{bocchichio,iwasakiward,mainimartinelli}.  They reported that the
crossing point $\beta_{ct}$ with this definition of the critical line is
located
in the region of strong coupling  on an $N_t=4$ lattice, {\it e.g.,}
$\beta_{ct}\approx 3.9-4.0$ for $N_f=2$.
For the phase diagram based on this result see ref.~\cite{wilsonreviews}.

This phase diagram, however, has an unsatisfactory feature.  It
has been observed\cite{qcdpaxwil,milcwil} that physical observables do
not
exhibit any singular behavior across the critical line in the high temperature
phase. This means that the region $K\geq K_c(\beta)$, usually thought
unphysical, is not distinct from the high temperature phase, being analytically
connected to it. Hence one can cross from the low- to the high-temperature
phase
through the part of the critical line below $\beta=\beta_{ct}$, which is not a
line of finite-temperature transition.

Clearly the phase diagram above does not capture the full
aspect of the phase structure.  Recent investigations indicate that a more
natural understanding of the phase structure is provided by a different view on
the critical line proposed by Aoki some time ago\cite{aoki}.
In the following we review the phase structure based on this
view.

Let us note that a slightly different phase structure has been discussed in
ref.~\cite{creutz}.  The phase structure for general values of $N_f$ up to
$N_f=300$ has also been examined recently\cite{qcdpaxlargenew}.

\subsection{Spontaneous breakdown of parity-flavor symmetry and massless pion}

In order to illustrate the basic idea, let us consider an effective sigma model
for lattice QCD with the Wilson quark action with $N_f=2$.  The effective
lagrangian may be written as
\be
{\cal L}_{eff}=(\nabla_\mu\vec\pi)^2+(\nabla_\mu\sigma)^2
+a\vec\pi^2+b\sigma^2+\cdots
\ee
where the coefficients $a$ and $b$ differ  reflecting explicit breaking of
chiral symmetry due to the Wilson term.  We know that the pion mass vanishes
as $a=m_\pi^2\propto K_c-K$ toward the critical line, while $\sigma$ stays
massive, {\it i.e.,} $b=m_\sigma^2>0$ at $K\approx K_c$.  If $K$ increases
beyond $K_c$, the coefficient
$a$ becomes negative.  Hence we expect the pion field to develop a vacuum
expectation value $\langle\vec\pi\rangle\ne 0$.  The condensate spontaneously
breaks parity and flavor symmetry.

Let us note that pion is not the Nambu-Goldstone boson of
spontaneously broken chiral symmetry in this view. Instead it represents the
massless mode of a parity-flavor breaking second-order phase transition which
takes place at $K=K_c$.  We expect it to become the Nambu-Goldstone boson
of chiral symmetry in the continuum limit, however, as chiral symmetry
breaking effects disappear in this limit.

The idea above has been explicitly tested for the two-dimensional
Gross-Neveu model formulated with the Wilson action\cite{aoki}.  An analytic
solution in the large $N$ limit shows spontaneous breakdown of parity for
$K\geq K_c(\beta)$.  Another important result of the solution is that
the critical line forms three spikes, which reach the weak-coupling limit $g=0$
at $1/2K=+2, 0, -2$.  This structure arises from the fact that the doublers at
the conventional continuum limit
$(g,1/2K)=(0,2)$ become physical massless modes at $1/2K=0$ and $-2$.

A close similarity of the Gross-Neveu model and QCD regarding the asymptotic
freedom and chiral symmetry aspects leads one to expect a similar phase
structure
for the case of QCD except that the critical line will form five spikes
reaching the continuum limit because of difference in dimensions\cite{aoki}.
Evidence supporting such a phase structure is summarized
in ref.~\cite{aokiyamagata}.

\subsection{Finite-temperature phase structure}

For a finite temporal lattice size $N_t$ corresponding to a finite temperature,
the above consideration can be naturally extend by defining the critical
line as the line of vanishing pion screening mass determined from the pion
propagator for large spatial separations.

\begin{figure}[t]
\centerline{\epsfxsize=74mm \epsfbox{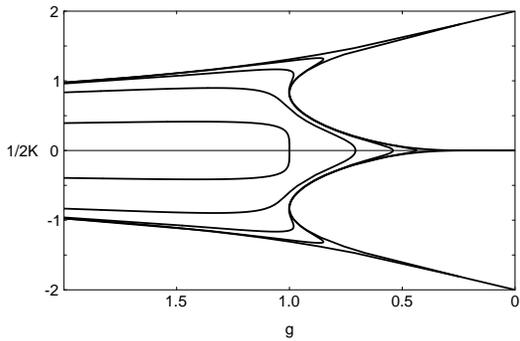}}
\vspace{-8mm}
\caption{Critical line in $(g,1/2K)$ plane for the two-dimensional Gross-Neveu
model for the temporal size $N_t=\infty, 16,8,4,2$ (from outside to
inside)
\protect\cite{auu}.}
\vspace*{-8mm}
\label{fig:fig6}
\end{figure}

In Fig.~\ref{fig:fig6} the critical line for the two-dimensional
Gross-Neveu model calculated in the large
$N$ limit is plotted for
$N_t=\infty, 16, 8, 4, 2$ starting from the outermost curve and
moving toward inside. The result shows that the location of the critical line
as
defined above depends on $N_t$.  Another important point is that the spikes
formed by the critical line moves away from the weak-coupling limit as $N_t$
decreases.

Simulations to examine if lattice QCD has a similar structure of the critical
line at finite temperatures have been made recently for
the case of $N_f=2$\cite{auu} and 4\cite{akuu} on an $8^3\times 4$
lattice.  The results are summarized as follows: (i) For both systems the
conventional critical line turns back toward strong coupling forming a cusp,
whose tip is located at $\beta\approx 4.0$ for $N_f=2$ and $\beta\approx 1.8$
for
$N_f=4$.  The cusp represents one of five cusps expected for lattice
QCD.\linebreak (ii) Parity and flavor symmetry are spontaneously broken inside
the cusp.  Simulations have been made for the $N_f=2$
system with an external field term
$\delta S_W=2KH\sum_n\overline{\psi}_ni\gamma_5\tau_3\psi_n$ added to the
action.   Results provide evidence for the behavior $\lim_{H\to
0}\langle\overline{\psi}\gamma_5\tau_3\psi\rangle\ne 0$ of the
parity-flavor order parameter and  vanishing of $\pi^\pm$ mass
$\lim_{H\to 0}m_{\pi^\pm}=0$ expected inside the cusp\cite{akuu}.

Concerning the relation between the thermal line and
the critical line, we recall that the pion mass vanishes all along the critical
line. This suggests that the region close to the critical line is in the cold
phase even after the critical line turns back toward strong coupling, and
hence the thermal line cannot cross the critical line.  Since numerical
estimates show that the thermal line comes close to the turning point of the
cusp, the natural possibility is that the thermal line runs past the tip of the
cusp and continues toward larger values of $K$.  Results of measurement of
thermodynamic quantities provide support of this view\cite{auu}, although the
possibility that the thermal line touches the critical line at a
point\cite{qcdpaxwil} cannot be excluded.

\begin{figure}[t]
\centerline{\epsfxsize=74mm \epsfbox{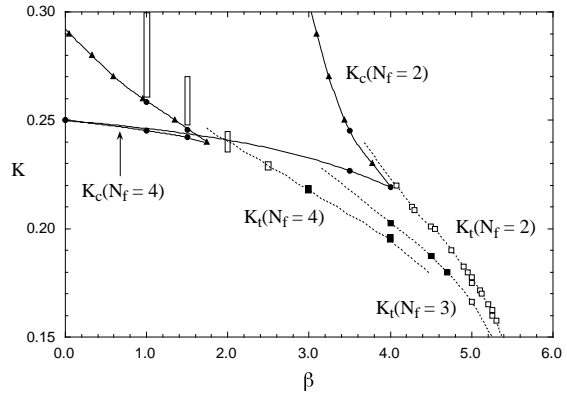}}
\vspace{-8mm}
\caption{Phase diagram for $N_f=2, 3, 4$ on an $N_t=4$
lattice.}
\vspace*{-8mm}
\label{fig:fig7}
\end{figure}

In Fig.~\ref{fig:fig7} we summarize presently available results for the phase
structure on an $N_t=4$ lattice for $N_f=2, 3$ and 4.  Solid lines represent
the critical line estimated from the pion mass.  For the case of $N_f=3$
results confirming the cusp structure are not yet available.  Open and solid
squares show simulation results for the location of the thermal line.
Dotted lines are smooth interpolation, extended beyond the cusp following the
discussion of the previous paragraph. The difference of open and solid squares
is discussed in Sec.~\ref{sec:order} below.

Let us remark that the conventional zero temperature critical line runs
close to the lower part of the finite-temperature critical line in
Fig.~\ref{fig:fig7} and continues toward weak coupling.  The thermal line
therefore has to cross the zero-temperature critical line.  This represents the
crossing point $\beta_{ct}$ reported by the QCDPAX
Collaboration\cite{qcdpaxwil}.

We emphasize, however, that the zero-temperature critical line does not
represent
a line of singularity of the finite-temperature partition function.  This
naturally explains the absence of singular behavior of observables across the
zero-temperature critical line mentioned in Sec.~\ref{sec:wilsonone}.

\subsection{$N_f$ dependence of order of transition}
\label{sec:order}

The sigma model analysis in the continuum\cite{sigmamodel}
suggests that the chiral phase transition is of second order for
$N_f=2$ and of first order for
$N_f\geq 3$. Indeed strong first-order signals have been observed for the case
of $N_f=3$\cite{qcdpaxwil} and 4\cite{akuu} away from the critical line, as
shown by solid squares in Fig.~\ref{fig:fig7}, in contrast to a crossover
behavior represented by open squares seen for $N_f=2$. However, the first-order
transition for $N_f=4$ weakens closer to the critical line, apparently turning
into a smooth crossover before reaching the region around the cusp of the
critical line as indicated by open rectangles\cite{akuu}. While parallel data
are
not yet available for
$N_f=3$, results of the QCDPAX Collaboration\cite{qcdpaxwil} also appears to
indicate a weakening of the first-order transition.

A possible reason for this unexpected behavior is that breaking of chiral
symmetry due to the Wilson term, which becomes stronger as $\beta$ decreases
along the thermal line, smoothens the first-order transition.  Another
possibility is that the first-order transition for $N_f=3$ and 4 observed so
far
is a lattice artifact sharing its origin with the sharpening of the crossover
at
$\beta\approx 5.0$ found by the MILC Collaboration for
$N_f=2$\cite{milcwil}.  Some support for this interpretation is given by a
recent study of the QCDPAX Collaboration for the $N_f=3$ system with an
improved
gauge action\cite{qcdpaximpthree}.  So far they have not found clear
first-order
signals in the region where the plaquette action shows a clear first-order
behavior.

In either case, if chiral transition in the continuum is indeed of first order
for $N_f=3$ and 4,  it will emerge only when the cusp moves sufficiently toward
weak-coupling with an increase of the temporal size $N_t$.

\subsection{Continuum limit}

We expect the cusp of the finite-temperature critical line
to grow toward weak coupling as $N_t$ increases.  In the limit $N_t=\infty$ it
should converge to the zero-temperature critical line which reaches
$\beta=\infty$.  Since the thermal line is located on the weak-coupling side of
the cusp for a finite $N_t$, it will be pinched by the tip of the cusp at
$(\beta, K)=(\infty, 1/8)$ as $N_t\to\infty$.  We expect chiral phase
transition
in the continuum to emerge in this limit.  In order to
extract  continuum properties of the chiral transition, we then need a
systematic study of thermodynamic quantities in the neighborhood of the
thermal line when it runs close to the tip of the cusp as a function of $N_t$.

Simulations, however, indicate that the cusp moves only very slowly
as $N_t$ increases.  For the $N_f=2$ case, current estimates of the position
of the tip of the cusp is
$\beta\approx 4.0$ for $N_t=4$\cite{auu}, $4.0-4.2$ for
$N_t=6$\cite{qcdpaxwil},
$4.2-4.3$ for
$N_t=8$\cite{akuu} and $4.5-5.0$ even for $N_t=18$\cite{qcdpaxwil}.  A
recent work also reports an absence of parity-broken phase above $\beta=5.0$ on
symmetric lattices up to the size $10^4$\cite{bitar}.  For $N_f=4$ the values
are
even lower:$\beta\approx 1.80$ for $N_t=4$ and $2.2-2.3$ for
$N_t=8$\cite{akuu}.
These estimates indicate that a very large temporal size will be  needed for
the cusp to move into the scaling region ({\it e.g.,} $\beta\geq 5.5$ for
$N_f=2$) as long as one employs the Wilson quark action together with
the plaquette action for the gauge part.

We emphasize that this result has an important implication also for spectrum
calculations at zero temperature.  Since the location of the cusp is determined
by the smaller of the spatial and temporal size, the critical line will
be shifted or may even be absent unless lattice size is taken
sufficiently large.  Therefore hadron masses calculated on a lattice of small
spatial size and extrapolated toward the position of the critical line might
involve significant systematic uncertainties.

\subsection{Studies with improved actions}

The problems discussed above indicate the presence of sizable cutoff
effects when the Wilson quark action is used in conjunction with the plaquette
action.  A way to alleviate this problem is to employ improved actions.  This
approach has been pursued by the QCDPAX
Collaboration\cite{qcdpaximp,qcdpaximpthree}, replacing the plaquette action
with an improved gauge action RG(1,2)\cite{iwasakirg}.  This year the MILC
Collaboration reported simulations with the action  SLW$_{tadpole}$
for the gauge part and the tadpole-improved clover action for the quark
part\cite{milcclover}.  Results with the tree-level clover action keeping
the plaquette action are also available\cite{auuclover}.  Thus there are data
for
four types of action combinations, unimproved and improved both for the gauge
and quark actions, to make a comparative study of improvement.

An indication from such a comparison is that improving the
gauge action substantially removes cutoff effects.  An inflection of the
critical line seen for the plaquette action at $\beta\approx 4-5$ becomes
absent
with improvement of the gauge action\cite{qcdpaximp}, while it still
seems to remains if only the Wilson quark action is replaced by the clover
action\cite{auuclover}. Also an intermediate sharpening of the thermal
transition seen for the plaquette action at $\beta\approx 5.0$\cite{milcwil}
is not observed for improved actions\cite{qcdpaximp,milcclover}.

Another point to note is that the lattice spacing at the coupling constant
where the thermal line approaches the critical line has a similar value $m_\rho
a\approx 1$ on an $N_t=4$ lattice for all of the four action combinations.
This means that studies of physical quantities are needed to assess
reduction of cutoff effects with improved actions.  Interesting results have
already been obtained for scaling of the chiral order
parameter\cite{qcdpaximp,qcdpaxsusceptibility}, and work with the critical
temperature is being pursued\cite{qcdpaximp,milcclover}.

\section{Results in finite density studies}

It has long been known that the quenched approximation breaks down for a
non-zero
quark chemical potential $\mu$ in that a transition takes place at $\mu\approx
m_\pi/2$ rather than at $\mu\approx m_N/3$\cite{barbourquenched,lombardo}.
While the importance of the phase of the quark determinant has been made clear,
the mechanism how the quenched approximation breaks has not been fully
explained.

Recently Stephanov\cite{stephanov}, employing a random matrix model of the
quark determinant\cite{randommatrix} and a replica formulation of quenched
approximation,
traced back the failure of the quenched approximation to the
non-uniformity of the limit of the replica number $n\to 0$ for
$\mu=0$ and $\mu>0$.  He has also shown that the quenched
approximation is valid for the theory in which  a quark $\chi$ in
the conjugate representation is added to each quark $q$.
Formation  of a condensate $\langle\overline{\chi}q\rangle$ having a
unit baryon number and a mass $m_{\overline{\chi}q}\approx m_\pi$ in
such a theory explains the occurance of transition at $\mu\approx m_\pi/2$.

Barbour and collaborators reported new results in full
QCD simulations\cite{morrison}.  With the method of fugacity
expansion\cite{fugacityexpansion} runs were carried out for four flavors
of quarks on $6^4$ and $8^4$ lattices at $\beta=5.1$ with the Kogut-Susskind
quark action.  They found an onset of non-zero baryon number at a small value
of
$\mu$, {\it e.g.}, $\mu_c\approx 0.1$ at $m_q=0.01$. For comparison the $MT_c$
Collaboration reported $m_N=1.10(6)$ and $m_\pi=0.290(6)$ at a slightly larger
coupling of $\beta=5.15$ at $m_q=0.01$\cite{mtc}.

It is not yet clear if these results mean that an early onset of transition
$\mu_c\approx m_\pi/2$ also holds for full QCD or reflect computational
problems of the method employed for the simulation.

\section{Conclusions}

Much work has been made in finite temperature studies of lattice QCD
encompassing a number of subjects during the last year.

Tests of improved actions made for the pure gauge system indicate a
possibility that accurate results for thermodynamics in the continuum may be
obtained with simulations carried out with a moderately large temporal size.

In full QCD studies much progress has been made in understanding the phase
structure for the Wilson quark action.  On the other hand, new
problems have also been encountered, making it necessary to reexamine
conclusions reached in previous studies.  These are the unexpected values of
exponents for $N_f=2$ found in scaling studies of susceptibilities with the
Kogut-Susskind quark action, and the flavor dependence of order of chiral
transition with the Wilson quark action.  Elucidating these problems is
important for reaching an understanding of the nature of chiral phase
transition, which is consistent between the Kogut-Susskind and Wilson quark
actions.

Some progress has been made in QCD at finite density.  A puzzling result
reported from the latest simulation shows, however, that we are still far
from understanding this difficult subject.

In closing we point out that most work in full QCD during the past
several years have concentrated on the case of $N_f$ degenerate quarks,
especially for $N_f=2$.  While a variety of basic problems we have encountered
for this case has to be clarified with further work, we should also recall that
nature corresponds to the case of $N_f=2+1$ with a heavier strange quark.  A
delicate change of phase that might possibly result from its presence, as
suggested in the continuum sigma model analysis\cite{sigmamodel}, makes it
important to enlarge
previous studies\cite{columbianf2,kogutsinclair,qcdpaxwil} into a systematic
effort in this direction.

\section*{Acknowledgements}

I would like to thank S. Aoki, F. Beinlich, T. Blum, W. Bock, G. Boyd,
M. Creutz, N. Christ, T. DeGrand, C. DeTar, M. Fukugita, Y. Iwasaki, K. Kanaya,
T. Kaneko, F. Karsch, E. Laermann, P. Mackenzie, M. Okawa,  D. Toussaint, M.
Wingate and Y. Yoshi\'e for communicating their results and for discussions.  I
would also like to thank S. Aoki, M. Fukugita, Y. Iwasaki, K. Kanaya and M.
Okawa for comments on the manuscript.  This work is supported in part by the
Grant-in-Aid of the Ministry of Education, Science and Culture (Nos. 04NP0801,
08640349).

\end{document}